\documentclass[sigconf]{acmart}

\AtBeginDocument{%
  }

\setcopyright{acmlicensed}
\acmPrice{15.00}
\acmDOI{10.1145/3540250.3559082}
\acmYear{2022}
\copyrightyear{2022}
\acmSubmissionID{fse22src-p84-p}
\acmISBN{978-1-4503-9413-0/22/11}
\acmConference[ESEC/FSE '22]{Proceedings of the 30th ACM Joint European Software Engineering Conference and Symposium on the Foundations of Software Engineering}{November 14--18, 2022}{Singapore, Singapore}
\acmBooktitle{Proceedings of the 30th ACM Joint European Software Engineering Conference and Symposium on the Foundations of Software Engineering (ESEC/FSE '22), November 14--18, 2022, Singapore, Singapore}

\begin{document}

\title{CheapET-3: Cost-Efficient Use of Remote DNN Models}

\author{Michael Weiss}
\email{michael.weiss@usi.ch}
\orcid{0000-0002-8944-389X}
\affiliation{%
  \institution{Università della Svizzera italiana}
  \city{Lugano}
  \country{Switzerland}
  \postcode{6900}
}

\begin{abstract}
 On complex problems, state of the art prediction accuracy of Deep Neural Networks (DNN) can be achieved using very large-scale models, consisting of billions of parameters. %
 Such models can only be run on dedicated servers, typically provided by a 3\textsuperscript{rd} party service, which leads to a substantial monetary cost for every prediction.
 We propose a new software architecture for client-side applications, where a small local DNN is used alongside a remote large-scale model, aiming to make easy predictions locally at negligible monetary cost, while still leveraging the benefits of a large model for challenging inputs. 
 In a proof of concept we reduce prediction cost by up to 50\% without negatively impacting system accuracy.
\end{abstract}



\maketitle

\section{Introduction}
Advances in machine learning showed a clear trend towards building ever larger Deep Neural Networks (DNNs):
AlexNet~\cite{krizhevsky2012alexnet}, the highly influential imagenet model released in 2012 set a milestone, with its then considered huge scale  60M parameters. 
Now, some models exceed even the trillion parameter threshold~\cite{fedus2021switch}.
Such huge models cannot be executed on resource constrained devices and environments such as microprocessors, mobile devices or in web browsers, but are instead hosted in server centers with  specialized hardware, causing substantial financial cost. 
E.g., a single request to the popular GPT-3 models~\cite{brown2020GPT3} is billed  up to \$0.48\footnote{4000 tokens on a fine-tuned davinci-model (0.12\$ per 1000 tokens) in July 2022. See \url{beta.openai.com}.}. 

In this research abstract, we present a software architecture designed to reduce the monetary costs of using large-scale DNNs.
We show a proof of concept (POC) of our approach using the the Imdb sentiment classification benchmark~\cite{maas2011IMDB}, with 25'000 training samples and 2000 randomly chosen test samples,
where we achieve a similar prediction performance to using pure GPT-3 predictions at only half the prediction cost.

\section{Proposed Architecture}
\begin{figure}
    \centering
    \includegraphics[width=0.4\textwidth]{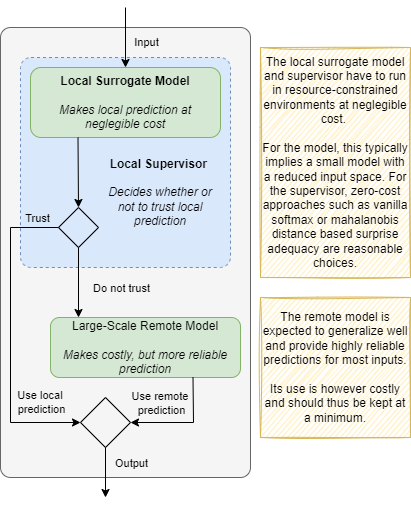}
    \caption{Proposed Architecture}
    \label{fig:architecture}
\end{figure}
Our proposed architecture is shown in~\autoref{fig:architecture}:
Instead of directly passing every input to the remote DNN, each input is first given to a small-scale and less accurate local  \emph{surrogate} model, s.t. \emph{easy, low uncertainty} predictions are made locally at negligible cost. 

A supervisor~\cite{Weiss2021FailSafe, Henriksson2019, Henriksson2019a, Hussain2022, xiao2021self, Weiss2021-SA, Ferreira2021, Stocco2020} is employed to detect inputs for which  local  predictions are confident enough to be trusted.
Untrusted predictions are forwarded to the large-scale remote model for a more reliable prediction.
Both local model and supervisor should be designed or chosen to account for the resource constraints:
Local models can e.g. use a compressed input space (e.g. small vocabulary sizes in NLP problems) and a reduced number of layers. 
For supervision, a wide range of techniques exist, which we compared in our previous work~\cite{Weiss2022SimpleTechniques, Weiss2021FailSafe, Weiss2021UncertaintyWizard, Weiss2021-SA, weiss2022ambiguity}.
We identified simple softmax-based supervisors, such as Vanilla Softmax (SM)~\cite{Hendrycks2016} (for classification problems) and Mahalanobis-distance based Surprise Adequacy (MDSA)~\cite{Kim2020MDSA} (for general problems) as efficient supervisors, which can be evaluated at negligible prediction-time cost~\cite{Weiss2022SimpleTechniques}:
The former is simply using the predicted softmax likelihood as a confidence score, where the latter measures uncertainty as the mahalanobis-distance~\cite{Mahalanobis1936} between a specific layer's activations for the observed input and the training set.

\section{Proof of Concept}
We evaluate our design using a small transformer~\cite{vaswani2017attention}, 
trained with a small vocabulary of 2000 tokens, as local surrogate model; the \textsc{text-curie-001} GPT-3 model~\cite{brown2020GPT3} (estimate 13 billion params~\cite{GP3ParamEstimate}, based on vocabulary with size 50257) as remote model, and independently evaluate both SM and MDSA as supervisor, using three different test sets (nominal, corrupted~\cite{Weiss2022SimpleTechniques} and partially-corrupted).
We then measure the resulting overall system accuracy, i.e., the accuracy given the local predictions for inputs where the supervisor trusted the local predictions, and the remote predictions for all other inputs.
In practice, the threshold as to when a supervisor trusts a local prediction could be continuously adapted to reach a target performance/cost ratio, hence we report results for a flexible percentage of inputs forwarded to GPT3.

\begin{figure}
    \centering
    \includegraphics[width=0.48\textwidth]{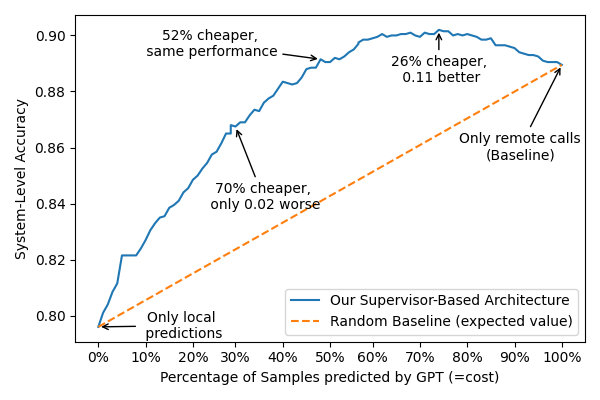}
    \caption{Performance of our architecture (nominal data, SM supervisor)}
    \label{fig:results}
\end{figure}
Results are shown in \autoref{fig:results}.
Due to space constraints, we only discuss the results for the nominal test set and the SM supervisor\footnote{The other results are similar, but with lower accuracy (i.e., the curve is shifted downward) for corrupted and partially-corrupted inputs and with slightly worse  (i.e., less curved) results for MDSA.}.
Clearly, our architecture allows for major cost savings, while maintaining a high prediction reliability:
In fact, by only making a prediction on 48\% of inputs, thus saving more than half of the GPT-3 cost, we achieve the same accuracy as if all predictions were made by GPT-3. 
Being even more cost-reduction oriented, with a 70\% cost saving accuracy decreases by only 0.02, thus still being much better than when using only the local model. 
Interestingly, we found that thanks to their complementary predictive capabilities, there is a combination  of  local and remote model that can lead to an overall system performance higher than the standalone, expensive remote model:
When sending 74\% of the inputs to GPT-3, accuracy increases by 0.11 while still saving 26\% on cost. 

\section{Conclusion \& Future work}
Our POC  shows the capability of our architecture to save monetary costs while only marginally -- if at all --  impacting the overall system performance. 
The architecture seamlessly generalizes beyond our POC to other classification and regression problems. It is also likely to lead to lower energy consumption and, on average, faster response time.
As future work, we  plan to evaluate this architecture on  different domains.%

\begin{acks}
Many thanks to Paolo Tonella and Andrea Stocco for their helpful advice. 
This work was partially supported by the H2020 project PRECRIME,
funded under the ERC Advanced Grant 2017 Program (ERC Grant Agreement n. 787703).
\end{acks}

\bibliographystyle{ACM-Reference-Format}
\bibliography{main}

\end{document}